\documentclass[a4paper, 11pt]{article}

\usepackage{graphicx,fancyhdr,amssymb}
\usepackage{amsmath,cite}
\usepackage{xcolor,booktabs}
\usepackage{soul}
\usepackage{url}
\usepackage{caption,authblk,cmbright}

\definecolor{c1}{RGB}{255,0,0}
\definecolor{c2}{RGB}{255,128,0}
\definecolor{c3}{RGB}{230,230,0}
\definecolor{c4}{RGB}{128,255,0}
\definecolor{c5}{RGB}{0,255,128}
\definecolor{c6}{RGB}{0,255,255}
\definecolor{c7}{RGB}{0,0,255}
\definecolor{c8}{RGB}{128,0,255}
\definecolor{c9}{RGB}{255,0,255}
\definecolor{c10}{RGB}{255,0,128}

\newcommand{\dd}{\ensuremath{{d}}}

\usepackage{hyperref}
\hypersetup{
	unicode=false,          
	pdftitle={Negative frequencies in pulse propagation equations and the double analytic signal},    
	pdfauthor={David Bermudez},     
	pdfsubject={Analogue Gravity},   
	pdfcreator={David Bermudez},   
	pdfproducer={David Bermudez}, 
	pdfkeywords={analogue gravity, Hawking radiation, optical fibres}, 
	pdfnewwindow=true,      
	colorlinks=true,       	
	linkcolor=magenta,      
	citecolor=blue, 			
	filecolor=red,      		
	urlcolor=red           	
}

\begin{document}

\lhead{Negative frequencies in pulse propagation and the DAS}
\rhead{Aguero-Santacruz, Bermudez}

\title{Negative frequencies in pulse propagation equations \\  and the double analytic signal}

\author{Raul Aguero-Santacruz\footnote{{\it email:} raul.aguero@cinvestav.mx} \ and David Bermudez\footnote{{\it email:} david.bermudez@cinvestav.mx}}
\affil{\textit{Department of Physics, Cinvestav, A.P. 14-740, 07000 Ciudad de M\'exico, Mexico}}

\vspace{10pt}

\date{}
\maketitle

\begin{abstract}

In recent years, the topic of negative frequencies has resurfaced in optics motivated by the optical analogue of Hawking radiation. We discuss the physical meaning of negative frequencies and the conditions under which they are relevant. We review how negative frequencies are treated in current pulse propagation models based on the electric field and the analytic signal. We focus on experimentally measured signals predicted by the conservation of negative comoving frequency in the nonlinear polarization terms to advance these concepts. We propose a new formalism called the double analytic signal which clearly separates negative frequencies from positive ones. Additionally, we reduce this new formalism to the analytic signal to prove their equivalence. Throughout the paper, we present numerical solutions of the unidirectional pulse propagation equation to illustrate the electric field, analytic signal, and double analytic signal formalisms and to highlight their differences.\\

\end{abstract}

\section{Introduction}
Light propagation in optical fibers is a major field of study, and the available literature is extensive \cite{golding1973nonlinear,tzortzakis2001self,karasawa2001comparison,Agrawal2008,Agrawal2013}. This situation creates a healthy field of research given the interplay between analytic predictions \cite{Kaup1990perturbation,Gorbach2007theory}, numerical solutions \cite{Potasek1986AnalyticAN,Roy2009effects}, and experimental verification \cite{Weiner2000FemtosecondPS,Dudley2009TenYO}. Optical fibers are mainly made of fused silica, which is a third-order nonlinear material \cite{Agrawal2013}. Many third-order nonlinear effects have been studied in detail, including four-wave mixing \cite{GarayPalmett2008, Garay2023}, dispersive waves \cite{hayes1973group,Akhmediev1995}, Raman shift \cite{Heller1982simple,rosenberg2020boosting}, and others.

Recently, there has been a focus on distinct and novel nonlinear effects that mix positive- and negative-frequency components of an optical field to generate new negative frequency signals, The negative frequency signals that have been experimentally verified so far are the negative-frequency resonant radiation (NRR) \cite{Rubino2012prl,Rubino2013,Drori2019} and the negative-frequency Hawking radiation (NHR) \cite{Drori2019,aguero2020hawking}. To understand these effects theoretically, an additional nonlinear term called conjugated Kerr effect or conjugated self-phase modulation (SPM*) must be added to the pulse propagation equations \cite{Conforti2013,Drori2019}. In some cases, the stated reason for omitting these terms is to discard duplicate data in the negative part of the spectrum resulting from a complex conjugated equation \cite{Agrawal2013,Boyd2008,newell2018nonlinear}; in other cases, this procedure is not even addressed \cite{mills2012nonlinear}. As a result of this thinking the SPM* term in the third-order nonlinearity is often discarded. These terms produce the negative frequency signals and are essential for energy conservation \cite{Conforti2013}.

It has taken many years to uncover these nonlinear terms because the conditions where negative frequencies are relevant have only recently become experimentally accessible. Technological advances in the form of ultra short pulses (on the order of a few optical cycles at $\sim10$ fs) \cite{Ma2019review} and the fabrication of highly-nonlinear fibers (as photonic-crystal fibers with a nonlinear coefficient $\gamma \sim10^{4}\; \text{W}^{-1} \text{km}^{-1} $) \cite{Russell2003photonic,Pandey2021} have led to a new regime of extreme nonlinear optics (XNLO) \cite{aguero2020hawking}, where negative frequencies are crucial. The reason is that the negative frequency branch of the dispersion relation is far from the positive branch. A short pulse in time has a wide frequency spectrum and together with high nonlinear interaction it can seed the negative branch.

The optical analogy of the event horizon \cite{Philbin2008,Bermudez2016pra,Drori2019,Felipe2022} has been the primary motivator to seriously consider the reality of negative frequencies in optics \cite{Biancalana2012}. The reason is that the fiber-optical analogue of Hawking radiation predicts the simultaneous emission of positive-frequency Hawking radiation and an entangled negative-frequency Hawking partner \cite{Philbin2008,Agullo2022}. Recent studies in optics on related topics have begun to include negative frequencies \cite{Robertson2019, Amiranashvili2022unusual}. Other physical systems that also exhibit horizon physics and negative-frequency signals include water waves \cite{Rousseaux2008,Rousseaux2010,Weinfurtner2011,Patrick2018}, polaritons in fluids of light \cite{Jacquet2022,Claude2023}, Bose-Einstein condensates \cite{Zapata2011,Munoz2019}, and graphene membranes \cite{Morresi2020}.

After Philbin et al. \cite{Philbin2008} proposed the optical analogue of Hawking radiation and its negative-frequency partner in 2008, the community quickly realized that they could more easily test the physical reality of negative frequencies by observing a different, more robust phenomenon known as NRR, i.e., the negative frequency equivalent of resonant radiation (RR), also called dispersive wave or Cherenkov radiation \cite{Akhmediev1995}. This phenomenon was experimentally verified by the work of Rubino et al. in 2012 \cite{Rubino2012prl,Rubino2013}. The negative-frequency Hawking partner of the analogue Hawking radiation (NHR) was finally measured by Drori et al. in 2019 \cite{Drori2019}, where they also measured the NRR. These experiments have motivated us to revisit the formalism to include the contribution of negative frequency signals. While similar ideas have gained wide acceptance in high-energy physics \cite{Thaller2013dirac}, there is still some resistance to considering negative frequencies in classical (i.e., non-quantum) optics \cite{Biancalana2012}.

\begin{figure}
	\centering
	\includegraphics[width=1.0\linewidth]{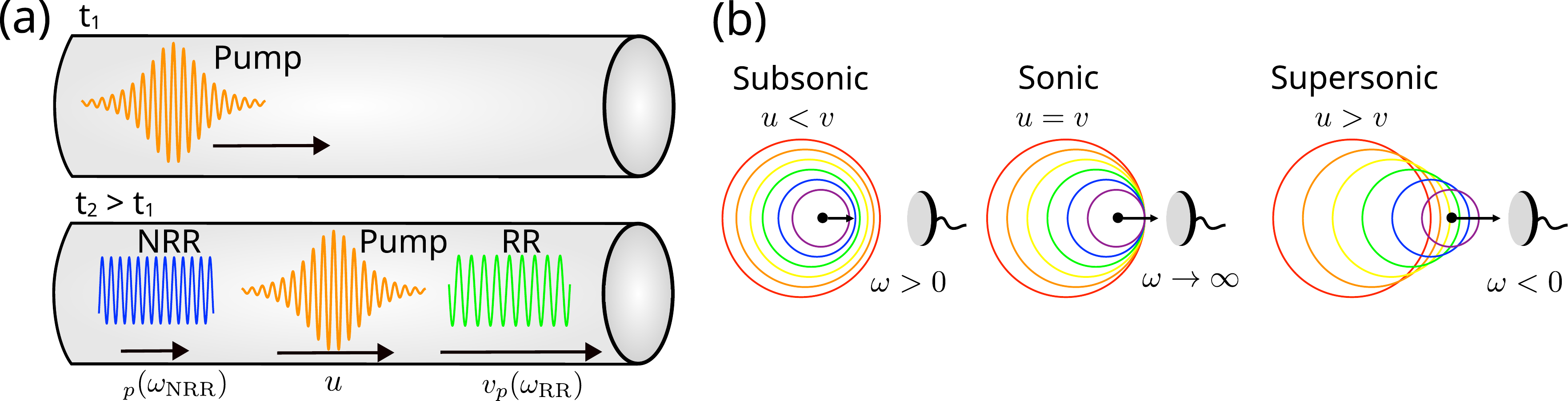}
	\caption{(a) Pulse propagation in an optical fiber that produces positive- (RR) and negative- (NRR) frequency signals. (b) Doppler effect for a sound emitter moving at different speeds $u$ with respect to the speed of sound $v$. The color indicates the time of emission. The supersonic case is interpreted as a negative-frequency signal because the waves arrive to the detector in reverse order from emission }
	\label{fibra}
\end{figure}

The usually discarded source term of the NRR signal is what we call the conjugated self-phase modulation term (SPM*), which was called the conjugated Kerr term in Ref. \cite{Conforti2013}\footnote{We believe that conjugated self-phase modulation (SPM*) is a more explicit name, especially considering the case of interacting pulses (not studied here), where the conjugated Kerr interaction includes several terms, including SPM*.}. Following the usual procedure \cite{Skryabin2005}, the resulting phase-matching condition for this new term predicts a measurable signal in positive frequencies, different from the regular self-phase modulation (SPM) term. Figure \ref{fibra} (a) shows the system under study, where a pump pulse propagates in an optical fiber. After some time, the nonlinear interaction of the pump pulse in the fiber produces both RR and NRR.

This work follows the correct mathematical formalism to derive the complete propagation equation for the electric field and the analytic signal, which includes the nonlinear interactions neccessary to generate the NRR signal. To support these ideas, we present a detailed description of the situation, analytical expressions, and numerical simulations. We emphasize the point of the derivation where negative frequencies are typically lost. Negative-frequency signals are weak and are usually overwhelmed by stronger signals even under favorable conditions in XNLO \cite{aguero2020hawking}. A way to separate the origin of such signals in theoretical and numerical studies can be of great value.

To facilitate the study of negative-frequency signals, we propose a new formalism that preserves the advantages of the analytic signal over the electric field and, more importantly, separates negative-frequency signals from positive ones. We base this formalism on a new mathematical object called the double analytic signal, which, unlike the analytic signal, includes negative frequencies. This formalism can become essential in analytical and numerical studies involving negative frequencies, and gives a new way to model known optical effects, leading, for example, to generalized versions of four-wave mixing and third-harmonic generation with negative frequencies. It can also be extended to other fields beyond optics, e.g., in the study of the Gross-Pitaevskii equation to model Bose-Einstein condensates \cite{Rogel2013gross,Bermudez2019}, where the mixing of positive- and negative-frequency signals is also studied \cite{Fabbri2021ramp,Kolobov2021observation}.

This manuscript is structured as follows: First, in Section \ref{secNeg}, we explain the physical reality of negative frequencies in general and in optics. Then, we study how to include negative frequencies in two formalisms: as the electric field in Section \ref{secEF} and as the analytic signal in Section \ref{secAS}. In Section \ref{secDAS}, we construct the new double analytic signal (DAS) formalism, and we reduce this formalism to be identical to the analytic signal. We present numerical solutions throughout the manuscript to illustrate our results. Finally, we write our conclusions in Section \ref{secCON}, where we compare the three formalisms.

\section{Physical reality of negative frequencies}\label{secNeg}
Negative frequencies can have physical meaning \cite{Biancalana2012}. We can show this fact with an example of sound waves in air suffering the Doppler effect. This example will then help us understand that the same physics is at play for light waves generated by a propagating pulse in a dielectric medium.

An emitter sends sound waves through the air with frequency $\omega'$ as measured from its own reference frame, i.e., its comoving frequency; a stationary detector measures a different frequency $\omega$, i.e., its laboratory frequency. The waves travel at the speed of sound $v$. (The prime refers to quantities in the reference frame comoving with the emitter.) The emitter now moves at speed $u$ toward a stationary detector. In the case of subsonic motion $u<v$, the frequency of the forward-moving waves increases, but their wavefronts still arrive at the detector in the order of emission: A measurement of positive frequency. In the sonic case $u=v$, we have a shock wave where a detector sees all the wavefronts arriving at once: A measurement of infinite frequency. Finally, for the supersonic case $u>v$, the first wavefront to arrive at the detector is the last one generated before the emitter passes the detector: This is a measurement of negative frequency, in the sense that the detector measures a time-reversed signal from the input source, i.e., the signals arrive in reverse order from the emission \cite{Rayleigh1896,Ahrens2008}. These three cases are illustrated in Figure \ref{fibra} (b) and can be obtained usign the non-relativistic Doppler formula:
\begin{equation}
	\omega'=\omega \frac{v-u}{v}.
\end{equation}
The detector in the laboratory frame does not distinguish between positive and negative frequencies and is only sensitive to the beat frequency of the signal $|\omega|$. Still, the comoving frequency can be negative in the supersonic case $u>v$.

We have a similar situation for light waves in a dielectric. A pulse of light travels at velocity $u$ and, due to nonlinear effects, emits light waves travelling at velocity $v_p(\omega)$, where $v_p(\omega)$ is the phase velocity of the medium and $\omega$ is the frequency of the emitted wave in the laboratory frame. Now we use the relativistic Doppler effect to study the system in the comoving frame and obtain the comoving frequency $\omega'$:
\begin{equation}\label{dooplerfreqs}
	\omega'= 
	\gamma \omega \frac{v_p(\omega) - u}{v_p(\omega)} ,
\end{equation}
where $\gamma$ is the relativistic Lorentz factor. First, the pump pulse with frequency $\omega_0$ enters the fiber and moves with its group velocity $u=v_g(\omega_0)$. The pump then emits two waves (RR and NRR) with different frequencies and velocities due to the dispersion relation of the fiber. This situation is shown in Figure \ref{fibra} (a). The equivalent of the speed of sound is the velocity of the emitted waves, given by the phase velocity $v_p(\omega)$. For RR, $u<v_p(\omega_\text{RR})$ as in the subsonic case and has a positive comoving frequency $\omega' > 0$. For NRR, $u>v_p(\omega_\text{NRR})$ as in the supersonic case and has a negative comoving frequency $\omega' < 0$. The transition point where the comoving frequency is $\omega' =0$ is equivalent to the sonic case $u=v_p(\omega_\text{ph})$. This condition is known as the phase horizon in horizon physics \cite{Bermudez2016pra} and is described by the frequency $\omega_\text{ph}$. These are the same three cases as in the example with sound waves in Figure \ref{fibra} (b).

\begin{figure}
	\centering
	\includegraphics[width=0.49\linewidth]{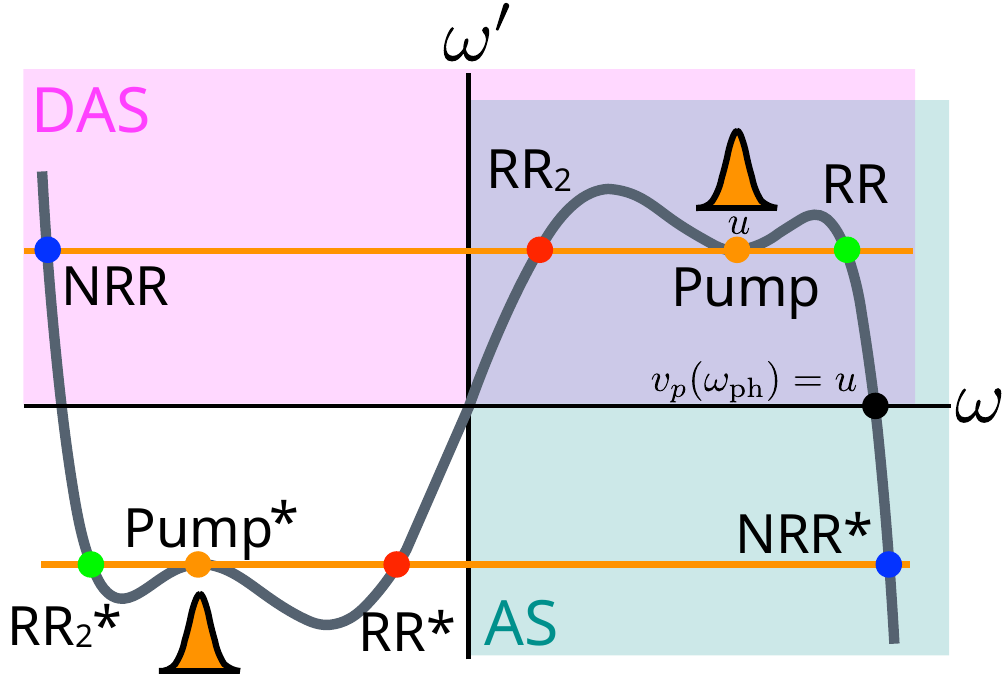}
	\caption{Regions of energy conservation for the analytic signal and DAS formalisms, where each signal has one point. The electric field includes all regions and two points per signal.}
	\label{qDAS}
\end{figure}

In both sound and light phenomena, the condition that defines negative frequencies is when waves reach the ``supersonic'' limit $u > v_p (\omega)$. In both cases we observe a shift from positive to negative comoving frequencies. This is because we use static detectors in the laboratory frame and $\omega\geq0$. Equivalently, if we set our detectors in the comoving frame then $\omega' \geq 0$ and the laboratory frequency could be negative. This means that the nature of negative frequencies depends on our point of view (more precisely, the detector's point of view). This fact is included in our choice of formalism to describe the system: the analytic signal keeps the laboratory frequencies positive, and our new formalism based on the double analytic signal keeps $\omega' \geq 0$. The electric field needs both positive and negative frequencies to be real (see Figure \ref{qDAS}). Note that for light, all velocities in both the laboratory and comoving frames are subluminal $u, v_p(\omega)<c$, even if waves satisfy the ``supersonic'' condition $u > v_p (\omega)$. For this reason, the word ``supersonic'' is sometimes used instead of ``superluminal'' in the study of optical event horizons \cite{Faccio2012}.

The comoving frequency $\omega'$ is an important quantity because it is conserved when the pump is slowly-varying or static (e.g., for a fundamental soliton). For this reason, it is simpler to describe the phase-matching condition as a conservation of $\omega'$, as we will see in Section \ref{secAS}. Figure \ref{qDAS} is a qualitative example of the dispersion relation in Eq. \eqref{dooplerfreqs}, where RR has a positive comoving frequency and NRR* has a negative comoving frequency, both in the region of positive laboratory frequency $\omega>0$. As we saw, the transition point happens at $\omega' = 0$ is given by the phase horizon frequency $\omega_{\text{ph}}$.

\section{Electric field formalism}\label{secEF}
When an optical pulse propagates in a nonlinear medium such as an optical fiber, many phenomena can occur that can be described by a formalism based on the electric field $E(z,t)$. These phenomena include the well-known resonant radiation (RR) (also called Cherenkov radiation or dispersive-wave radiation) \cite{hayes1973group,Akhmediev1995,Agrawal2013}, Raman-induced frequency shift \cite{Crawford2019,rosenberg2020boosting}, and supercontinuum generation \cite{Dudley2010}. It is possible to derive a propagation equation by manipulating Maxwell's equations and using several well-suited approximations, either for the electric field or its envelope \cite{Couairon2011}. Of the two variables, the envelope equations are the most commonly used equations due to their ease of numerical implementation.

For the simplest case of linearly polarized light in a defined transverse mode, we consider the electric field decomposition into an amplitude $A(z,t)$ times a carrier wave with frequency $\omega_0$ of the form
\begin{eqnarray}\label{carrier-envelope}
E(z,t) &=|A(z,t)| \cos[{\omega_0 t +\theta(z,t)}] = \frac{1}{2} \left[ A(z,t) e^{-i \omega_0 t} +A^*(z,t) e^{i \omega_0 t} \right],
\end{eqnarray}
where $A(z,t) =|A(z,t)| e^{i\theta(z,t)}$ is the envelope, $z$ is the propagation direction, and $t$ is the time variable. 
Equations with the electric field $E$ as its variable are carrier-resolved propagation equations. They include the unidirectional pulse propagation equation (UPPE), the forward Maxwell equation (FME), the forward wave equation (FWE), and the first-order propagation equation (FOP). On the other hand, envelope propagation equations include the forward envelope equation (FEE), the nonlinear envelope equation (NEE), and the nonlinear Schrödinger equation (NLSE) \cite{Kolesik2004,Couairon2011}.

One of the most important parts of any propagation model is the dispersion component. Either considering the full propagation constant $\beta(\omega)$ or neglecting higher-order terms. At least up to third-order dispersion $\beta_3$ is necessary to capture the physics to display resonant radiation phenomena. For ultra short pulses ($\sim10$ fs), we must consider the full dispersion model for the propagation equation $\beta(\omega)=\omega n(\omega)/c$, where $n(\omega)$ is the refractive index of the fiber \cite{Agrawal2013}. We will only consider a single propagation direction, neglecting any backward propagating waves. Additionally, we ignore the Raman shift and energy loss due to propagation. The UPPE of the electric field has the desired properties for use in XNLO: \begin{equation}\label{UPPE}
 i \partial_z E_\omega +\beta(\omega) E_\omega + \frac{\omega}{2cn(\omega)} P_{\text{NL},\omega} =0,
\end{equation}
where $E_\omega (z)= \int ^{\infty}_{-\infty} E(z,t) e^{i \omega t} \text{d}t$ is the Fourier transform of the electric field $E(z,t)$, and the factor $\omega/n(\omega)$ is the dispersion of the nonlinearity, which modulates the self-steepening effect \cite{Philbin2008,Agrawal2013}. The variable $E_\omega$ exists in the domain $\omega\in(-\infty,\infty)$ and is an even function due to $E(z,t)$ being real (since it is a physical quantity). The nonlinear term $P_\text{NL}= \chi^{(3)}E^3$ acts as the source of the new frequencies appearing in the spectrum during propagation. The paradigmatic NLSE is a particular case of the UPPE that includes only the second-order dispersion and one nonlinear term (SPM).

In the case of Eq. \eqref{UPPE}, there are two inconveniences: (i) no control over the individual nonlinear effects, and (ii) duplicated information since the negative part of the spectrum is symmetrical to the positive side. There is no distinction between the positive- and negative-frequency signals, as they appear on both sides of the spectrum. On the plus side, there is no possible confusion regarding additional simplifications on the negative frequencies. In the next section, we will see that oversimplifications can occur when switching to a different formalism based on the analytic signal.

\begin{figure}
	\centering
	\includegraphics[width=0.49\linewidth]{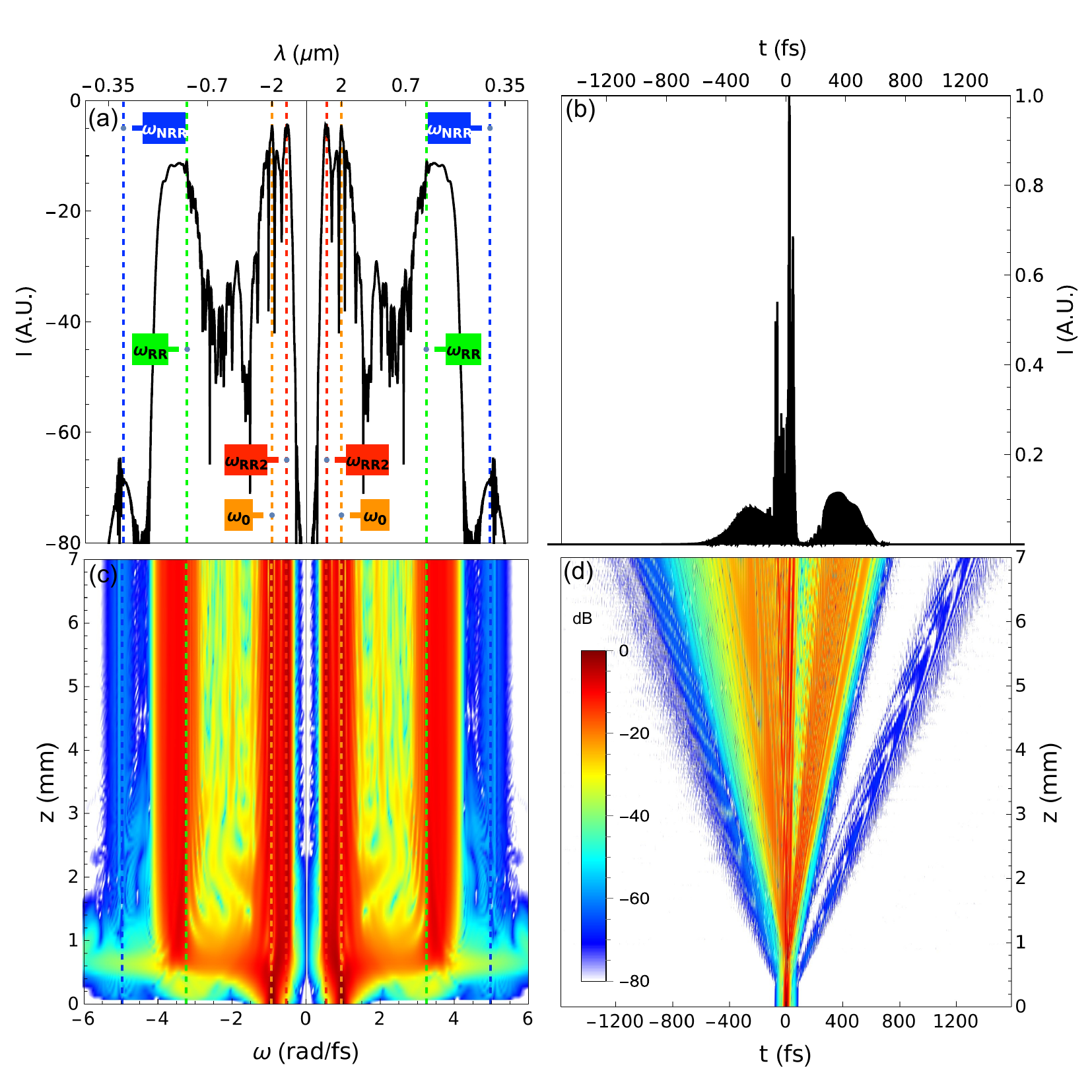}
	\caption{Electric field propagation, where (a) and (b) are the final state in the propagation, and (c) and (d) show the spectral and temporal evolution, respectively.}
	\label{EF}
\end{figure}

In Figure \ref{EF}, we show a solution of the UPPE for the electric field \eqref{UPPE} for a short soliton pulse $P_0 \text{sech}^2(\tau/\tau_0)$ with peak power $P_0=2\times 10^5$ W and duration $\tau_0 =7$ fs propagating in fused silica. We use the dispersion relation for fused silica from Ref. \cite{Kasap2006}, which will be discussed in the next section. Panels (a) and (c) show that the spectrum is the same for $\omega>0$ and $\omega < 0$. The colored vertical lines are the theoretical predictions for the newly generated signals, including two RRs and one NRR in both positive and negative laboratory frequencies. Panels (b) and (d) show the temporal evolution of the pulse. We provide more information on the numerical implementation in \ref{AppNum}.

\section{Analytic signal formalism}\label{secAS}
Mathematically, negative frequencies are embedded in the most common functions used to describe waves. Every real sine or cosine is composed of two complex exponentials with positive and negative frequencies $\pm\omega$, as shown in Euler's formula:
\begin{equation}\label{sinecosine}
	\sin{\omega t} = \frac{e^{i\omega t}-e^{-i\omega t}}{2i},\quad
	\cos{\omega t} = \frac{e^{i\omega t}+e^{-i\omega t}}{2}.
\end{equation}
A real signal $X(t)$ with spectrum $X_\omega$ satisfies $|X_\omega|=|X_{-\omega}|$, i.e., the spectrum has the same amount of information for both $\omega>0$ and $\omega<0$.

From Eq. \eqref{sinecosine} we deduce that the complex single-frequency plane wave $A e^{i\omega t}$ with positive frequency is simpler than the real double-frequency plane wave $A \sin{(\omega t)}$ with positive and negative frequencies \cite{smith2008mathematics}. In this sense, working with complex exponentials instead of trigonometric functions has two advantages: (i) the number of components is reduced by half, and (ii) analytical and numerical calculations are simpler. Thus, light---as a wave---can be described either by real-valued sines (or cosines) through the electric field or by complex-valued exponentials through the analytic signal \cite{Couairon2011}.

The decomposition in Eq. \eqref{carrier-envelope} is equivalent to expressing a real number $x$ in terms of its complex components $z$, such that $x=(z+z^*)/2$. In a similar decomposition, we expand a real function into an infinite sum of its complex Fourier components \cite{Amiranashvili2022}. The analytic signal $\mathcal{E}(z,t)$ was developed with this vision in mind \cite{smith2008mathematics} as a simplified complex version of the electric field. The analytic signal is defined as
\begin{equation}
\mathcal{E}(z,t) =  \frac{1}{\pi}\int^{\infty}_0 E_\omega (z) e^{-i\omega t} \text{d}\omega.
\end{equation}
Note that the domain of the integral is $\omega\in[0,\infty)$, and it is not simply the inverse Fourier transform of the electric field. Its Fourier transform $\mathcal{E}_\omega$ is
\begin{equation}
\mathcal{E}_\omega =\mathcal{F}^{}\{\mathcal{E}(t) \}=  \int^\infty_{-\infty} \mathcal{E}(t) e^{-i \omega t} \text{d} t,
\end{equation}
with the inverse
\begin{equation}
\mathcal{E}(t) =\mathcal{F}^{-1}\{  \mathcal{E}_\omega \}= \frac{1}{2\pi} \int^\infty_{-\infty} \mathcal{E}_\omega e^{i \omega t} \text{d} \omega,
\end{equation}
recall that $\mathcal{E}_\omega$ is zero for $\omega <0$. The analytic signal depends only on the positive-frequency half of the electric field, according to its definition. This new function is complex; its complex conjugated includes the negative-frequency half of the electric field. We can derive the relationship between the analytic signal and the electric field and between their respective Fourier transforms as
\begin{equation}\label{EFtoAS}
E = \frac{1}{2} [\mathcal{E} + \mathcal{E}^*], \quad E_\omega = \frac{1}{2} [\mathcal{E}_\omega + (\mathcal{E}^*)_{\omega}].
\end{equation}
An additional definition of the analytic signal in terms of the Hilbert transform $\mathcal{H}$ is
\begin{align}
\mathcal{{E}}(z,t) &= E(z,t)-i \mathcal{H}[E(z,t)],\\
\mathcal{H}[E(z,t)] &= \frac{1}{\pi} \mathcal{P} \int^{\infty}_{-\infty} \frac{E(z,t')}{(t-t')} \text{d} t',
\end{align}
where $\mathcal{P}$ is the principal value \cite{Conforti2013}. The Fourier transform of the analytic signal satisfies the relation $\mathcal{E}_\omega (z) = [1+\text{sgn} (\omega)] E_\omega (z)$ and $\mathcal{E}_{\omega=0} = E_{\omega =0}$.

The analytic signal is sometimes considered an envelope of the electric field. It has some of the properties of an envelope, but they are not strictly identical \cite{Couairon2011}. In particular, $A(z,t)$ in Eq. \eqref{carrier-envelope} includes positive and negative frequencies, while $\mathcal{E}(z,t)$ in Eq. \eqref{EFtoAS} only positive frequencies. We can define an envelope for the analytic signal $\mathcal{A} (z,t)$ as
\begin{equation}\label{ASenv}
	\mathcal{E}(z,t) = \mathcal{A} (z,t) e^{-i\omega t}.
\end{equation}
We split the frequency and energy content of the field into positive and negative halves. The analytic signal has advantages over the electric field in both theoretical and numerical implementations. Unlike the electric field envelope, the analytic signal does not require approximations in its derivation and does not define a carrier frequency $\omega_0$ \cite{Amiranashvili2022}. When using the envelope of the analytic signal instead of the analytic signal as the main variable, Eq. \eqref{ASenv} acts as the bridge connecting the validation of the theory presented for the analytic signal with envelope equations.

We always consider positive and negative components when dealing with the electric field. However, we end up with duplicated information because the positive-frequency part of the spectrum is symmetrical to the negative one. This electric field $E$ behavior can easily observed in numerical implementations, as shown in Figure \ref{EF}. The implied physical reality is that every measured signal has positive and negative frequencies simultaneously, even if it is impossible to distinguish them in the laboratory.

Considering the analytic signal as our variable, only positive frequencies are present and negative frequencies are discarded or filtered out. It is crucial to note that the analytic signal formalism leads to identical physical predictions as the electric field. However, there are some advantages of the analytic signal approach over the electric field: (i) more flexibility in controlling individual parts of the nonlinearity---and the underlying physics---by splitting it into three terms (THG, SPM, and SPM*), (ii) ease of work with exponential functions instead of trigonometric functions in theoretical and numerical implementations, and (iii) numerical implementations are faster due to the half length of the field array (even if it is complex). However, some disadvantages are that it is easier to erroneously discard the SPM* term and that negative-frequency signals still appear in the positive (conjugated) side of the spectrum.

Let us consider the transformation from $E$ to $\mathcal{E}$ in Eq. \eqref{EFtoAS} and apply it to the UPPE \eqref{UPPE}. We have a new equation in which both $\mathcal{E}$ and $\mathcal{E}^*$ appear simultaneously. The linear part has the same form, and the nonlinear part separates into four different terms:
\begin{equation}\label{NLterms}
E^3 = \mathcal{E}^3+3\mathcal{E}^2\mathcal{E}^*+3\mathcal{E}^{*2}\mathcal{E}+\mathcal{E}^{*3},
\end{equation}
the first two terms are the third-harmonic generation (THG) and the self-phase modulation (SPM). The last two are uncommon, but we can call them conjugated self-phase modulation (SPM*) and conjugated third-harmonic generation (THG*) \cite{Conforti2013}. The separation into four terms in the analytic signal formalism allows for more control of the nonlinear contribution of each term, making it easier to study individual effects.

We could write the nonlinear part as $E^3=\mathcal{E}^3+3\mathcal{E}^2\mathcal{E}^*+ \text{c.c.}$ and do the same for the linear terms. Then, the complete equation is
\begin{equation}\label{UPPEcomplete}
 i \partial_z \mathcal{E}_\omega +\beta(\omega) \mathcal{E}_\omega + \frac{\omega}{16cn(\omega)} \chi^{(3)} [\mathcal{E}^3+3\mathcal{E}^2\mathcal{E}^*]_\omega + \text{c.c.}=0.
\end{equation}
A similar derivation can be followed for the complex envelope of the electric field $A$ \cite{Boyd2008,mills2012nonlinear,Agrawal2008,newell2018nonlinear}. At this stage, we may consider discarding the c.c. terms as they appear to contain only duplicated information in the negative-frequency side of the spectrum. This assumption is partially incorrect, since the c.c. terms actually contain some information in the positive-frequency side, but this information is duplicated in the negative-frequency side. So we can keep all the information in two ways: including the negative frequencies without the c.c. terms, or including the c.c. terms and only the positive frequencies. The first scenario is possible for the envelope of the electric field $A$, but not for the analytic signal $\mathcal{E}$ nor for the envelope of the analytic signal $\mathcal{A}$, because the last two objects are defined only for positive frequencies. The resulting equation for the analytic signal is 
\begin{equation}\label{UPPEincomplete}
	i \partial_z \mathcal{E}_\omega +\beta(\omega) \mathcal{E}_\omega + \frac{\omega}{16cn(\omega)} \chi^{(3)} [\mathcal{E}^3+3\mathcal{E}^2\mathcal{E}^*]_\omega=0.
\end{equation}
This equation is ill-defined due to the appearance of negative frequencies in the nonlinear term since the analytic signal is defined for $\omega\in[0,\infty)$. Then we must keep the resulting field of the nonlinear part as positive and the SPM term in Eq. \eqref{UPPEincomplete} does not maintain this property. This problem does not appear in the usual derivations of this equation based on the envelope of the electric field $A$ as further over-constraining assumptions are used to justify this equation, as the ubiquitous slowly-varying envelope approximation (SVEA) that is used to consider only frequencies in the neighborhood of the positive carrier frequency and, therefore, to keep only the SPM term \cite{Boyd2008,mills2012nonlinear,Agrawal2008,newell2018nonlinear}. Furthermore, even if the information is there in principle, negative-frequencies are usually neglected in these derivations, as the negative-frequency matching does not commonly appear in those derivations.

If we filter the nonlinear part in \eqref{UPPEincomplete} to keep $\omega\in [0,\infty)$, then the resulting equation
\begin{equation}\label{UPPEill}
i \partial_z \mathcal{E}_\omega +\beta(\omega) \mathcal{E}_\omega + \frac{\omega}{16cn(\omega)} \chi^{(3)} [\mathcal{E}^3+3\mathcal{E}^2\mathcal{E}^*]_{\omega +}=0,
\end{equation}
is incomplete. Mathematically, a Heaviside or step function $\Theta(\omega)$ applies the positive filtering. However, the SPM term has some information in $\omega<0$. Then, filtering the SPM term deletes this information. In particular, Eq. \eqref{UPPEill} does not produce the NRR signal.

Instead, we should return to the complete Eq. \eqref{UPPEcomplete} and apply the positive filter \footnote{We can also use a negative filtering, but positive is most common.}. This filtering immediately discards terms such as $i\partial_z \mathcal{E}^{*}_\omega$ or $(\mathcal{E}^{*3})_\omega$, as they contain only negative frequencies because the analytic support of $\mathcal{E}^*$ is $\omega\in(-\infty,0]$. The terms $3(\mathcal{E}^2\mathcal{E}^*)_\omega$ or $3(\mathcal{E}\mathcal{E}^{*2})_\omega$ remain in the equation with a $+$ subscript indicating the positive filtering. These terms mix regular and conjugated analytic signals and contain positive and negative frequencies, since their analytical support is the domain $\omega\in(-\infty,\infty)$. With this filtering procedure, the correct propagation equation for the analytic signal is
\begin{equation}\label{UPPEcorrect}
	i \partial_z \mathcal{E}_\omega +\beta(\omega) \mathcal{E}_\omega + \frac{\omega}{16cn(\omega)} \chi^{(3)} [\mathcal{E}^3+3\mathcal{E}^2\mathcal{E}^*+3\mathcal{E}^{*2}\mathcal{E}]_{\omega +} =0.
\end{equation}
 The main differences between Eqs. \eqref{UPPEincomplete} and Eq. \eqref{UPPEcorrect} are the addition of the SPM* term $\mathcal{E}^{*2}\mathcal{E}$ and the filtering procedure, where $\mathcal{E}^3$ is unaffected, $\mathcal{E}^2\mathcal{E}^*$ and $\mathcal{E}^{*2}\mathcal{E}$ are partially filtered, and $\mathcal{E}^{*3}$ is completely eliminated. The additional SPM* term in the propagation equation acts as a new source that generates signals produced by the mixing of positive and negative frequencies, which have been observed both numerically and experimentally in the positive part of the spectrum. A signal produced by this term, called negative-frequency resonant radiation (NRR), has already been predicted and measured in the laboratory by Rubino et al. in 2012 \cite{Rubino2012prl} and later by Drori et al. in 2019 \cite{Drori2019}. A theoretical and numerical study of the NRR phenomenon was published by Conforti et al. in 2013 \cite{Conforti2013}. Therefore, it is clear that the correct view is frequency filtering Eq. \eqref{UPPEcomplete} before discarding the complex conjugated part in Eq. \eqref{UPPEincomplete}. It is worth noting that the signals produced by SPM* are only visible in XNLO conditions and, as such, do not commonly appear in fiber optical experiments The experimental conditions for the regime of XNLO include ultrashort pulses ($\sim10$ fs) \cite{Drori2019} and the use of highly nonlinear fibers (as the photonic-crystal fibers) with nonlinear coefficient $\gamma\sim10^4$ W$^{-1}$ km$^{-1}$ \cite{Rubino2012prl,Drori2019}.

Another way of solving solving the ill-definition in Eq. \eqref{UPPEincomplete} it to expand the domain of the field to include negative frequencies. This is the approach we follow with the introduction of a new formalism in Section \ref{secDAS}.

\begin{figure}
	\includegraphics[width=0.49\linewidth]{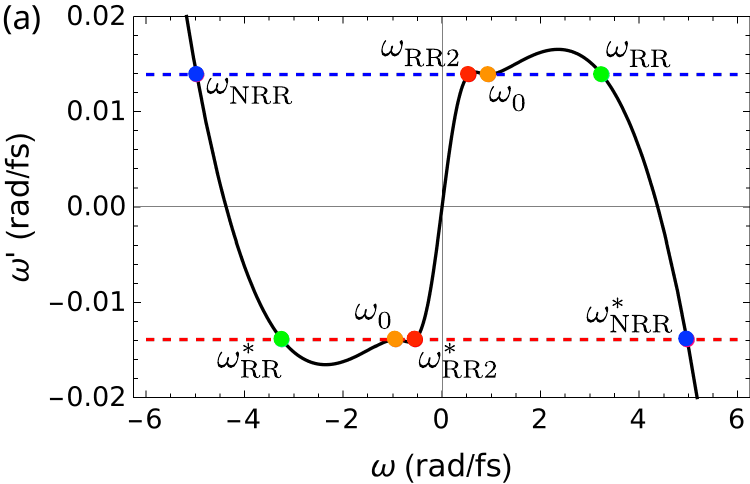}
	\includegraphics[width=0.49\linewidth]{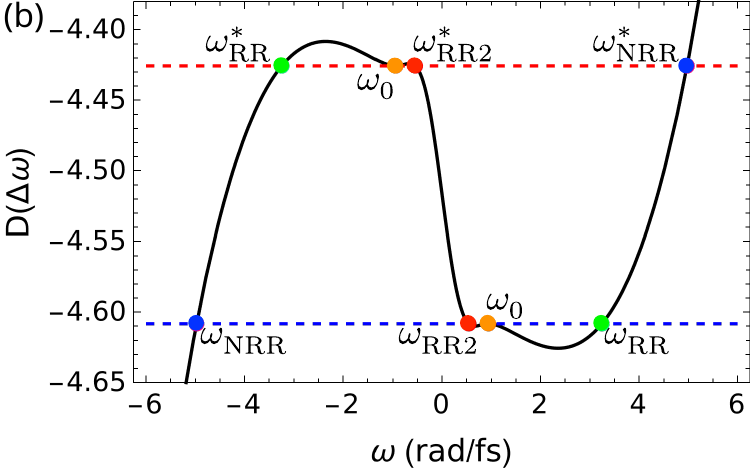}
	\caption{ (a): Dispersion relation $\omega^{\prime}(\omega)$ in the comoving frame for fused silica. The intersections with $\omega^{\prime}(\omega)=\omega^{\prime}(\omega_0)$ (blue) and $\omega^{\prime}(\omega)=-\omega^{\prime}(\omega_0)$ (red) give the resonant frequencies for SPM and SPM*, respectively. (b): Dispersion relation $D(\omega)$ in the laboratory frame and the corresponding conservation laws and resonant frequencies.}
	\label{f11}
\end{figure}

A common technique when working with propagation equations is to describe the dynamics in a comoving frame under the coordinate transformation $\zeta = z$ and $\tau = t-z/u$, where the frame velocity $u$ is chosen for convenience as the group velocity at the central frequency of the pump pulse. The equation is valid in any comoving frame, even if the pump has a varying group velocity. Indeed, this is the case in all numerical examples presented in this work. As seen in Section \ref{secNeg}, for this transformation we must also consider a comoving frequency $\omega'$ as the result of a relativistic Doppler shift, where its relation to the laboratory frequency is given by
\begin{equation}
 \omega'(\omega) = \omega-u \beta(\omega),
\end{equation}
which is another way of expressing the relation \eqref{dooplerfreqs} and where we have absorbed $\gamma$ in $\omega'$, as usual in optics. We can get the same information from the laboratory and comoving frame by using the propagation constant $\beta(\omega)$ or the comoving frequency $\omega' (\omega)$, respectively. However, in most scenarios involving an envelope equation and a change of coordinates, common simplifications lead to the use of
\begin{equation}
D(\Delta\omega)=D(\omega - \omega_0)=\beta(\omega-\omega_0)-\beta_1 (\omega - \omega_0) -\beta_0= \sum^{\infty}_{n=2} \beta_n (\omega- \omega_0)^n/n!
\end{equation}
for the laboratory system. There is no distinction between positive and negative comoving frequencies $\omega'$ in $D(\Delta \omega)$. A phase-matching condition for one formalism can be translated to all others, as shown in Figure \ref{f11}. In both cases, a procedure described in Ref. \cite{Skryabin2005} can be followed to obtain conditions represented as conservation laws of momentum in the laboratory frame and of energy in the comoving frame.

The conservation of comoving frequency leads to phase-matching of the RR process from the SPM term: $\omega'(\omega)=\omega'(\omega_0)$, and of the NRR process from the SPM* term: $\omega'(\omega)=-\omega' (\omega_0)$. Therefore, we can have negative comoving frequencies in the region of positive laboratory frequency. This matching is evident in Figure \ref{f11}, where the predicted signal has a positive laboratory frequency $\omega$, but a negative comoving frequency $\omega'$. The same predicted signals occur for $\omega>0$, but with opposite signs in the region $\omega<0$. This duplication of newly generated signals is what we typically encounter when using the electric field formalism.

In the analytic signal formalism, we are restricted to the $\omega>0$ region with no restriction on the sign of $\omega'$, covering quadrants I and IV in the dispersion relation, as shown in Figure \ref{qDAS}. For the electric field formalism, both $\omega$ and $\omega'$ values are unrestricted and cover all four quadrants. Consequently, the electric field contains duplicated information, since the propagation constant is an odd function \cite{Agrawal2008}. Both Eqs. \eqref{UPPE} and \eqref{UPPEcorrect} are connected by relation \eqref{EFtoAS} and describe the same phenomena for optical pulse propagation.

We note that although the appearance of negative frequencies can be seen as a consequence of a Doppler shift in frequency due to the change of reference frame, they are actually present in all frames. We introduce an asymmetry by considering only positive frequencies in the frame where the detector is stationary. This is because physical detectors respond to the beating of a signal. If the detector is in the laboratory frame, we consider only positive laboratory frequencies, and the comoving frequencies can be negative (as in the analytic signal formalism). On the other hand, if our detector is in the comoving frame, we must consider negative laboratory frequencies (as in the double analytic signal formalism). The inclusion of negative frequencies is unavoidable.

By keeping the SPM* term and the + filter in the analytic signal, we can obtain the same frequency profile in the positive-frequency spectrum as in the electric field, where the same signals are present in both scenarios, including those with negative comoving frequency (as NRR). Formally, only the THG* term should be eliminated since its spectral content is composed exclusively of negative frequencies.

\begin{figure}
\centering
\includegraphics[width=0.49\linewidth]{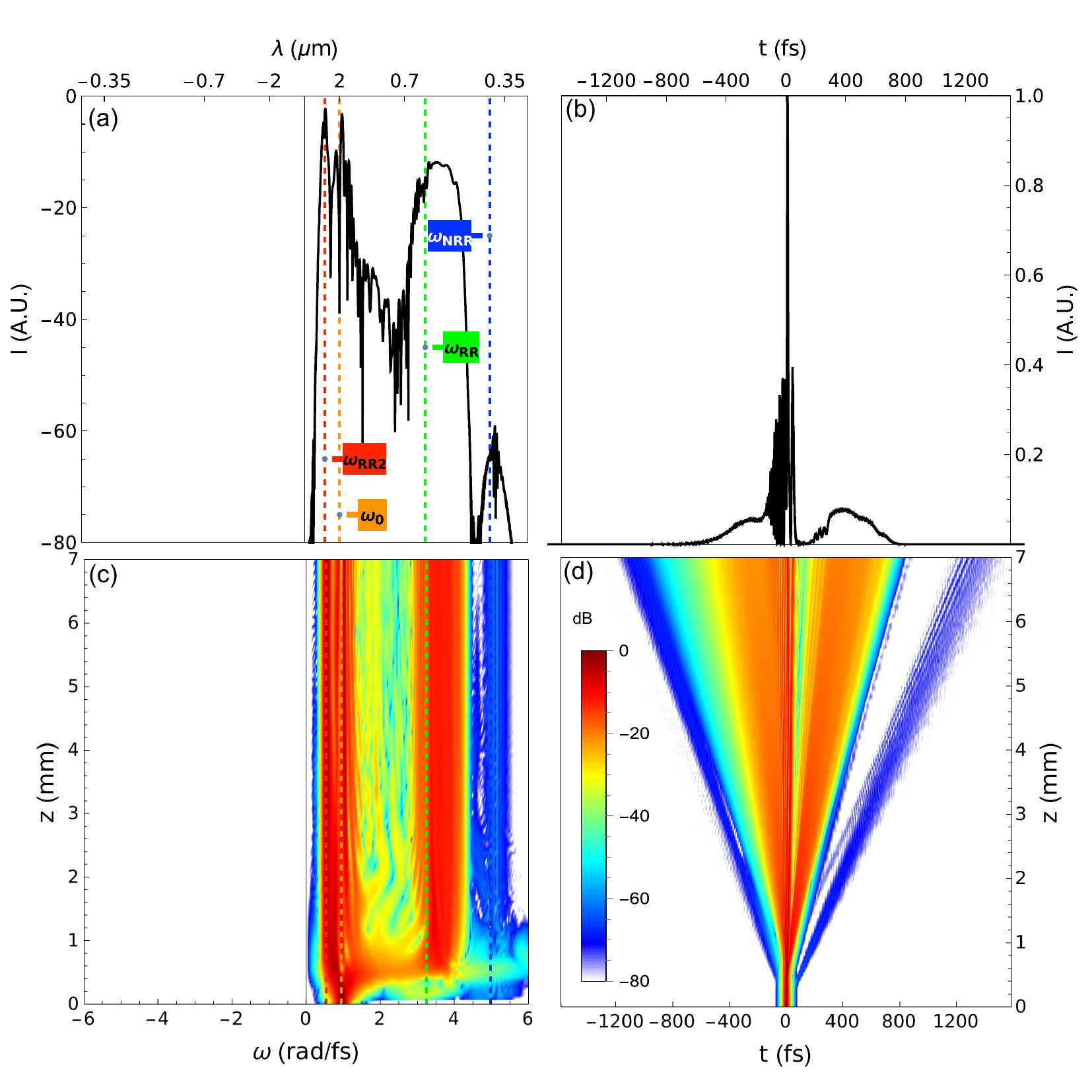}
\includegraphics[width=0.49\linewidth]{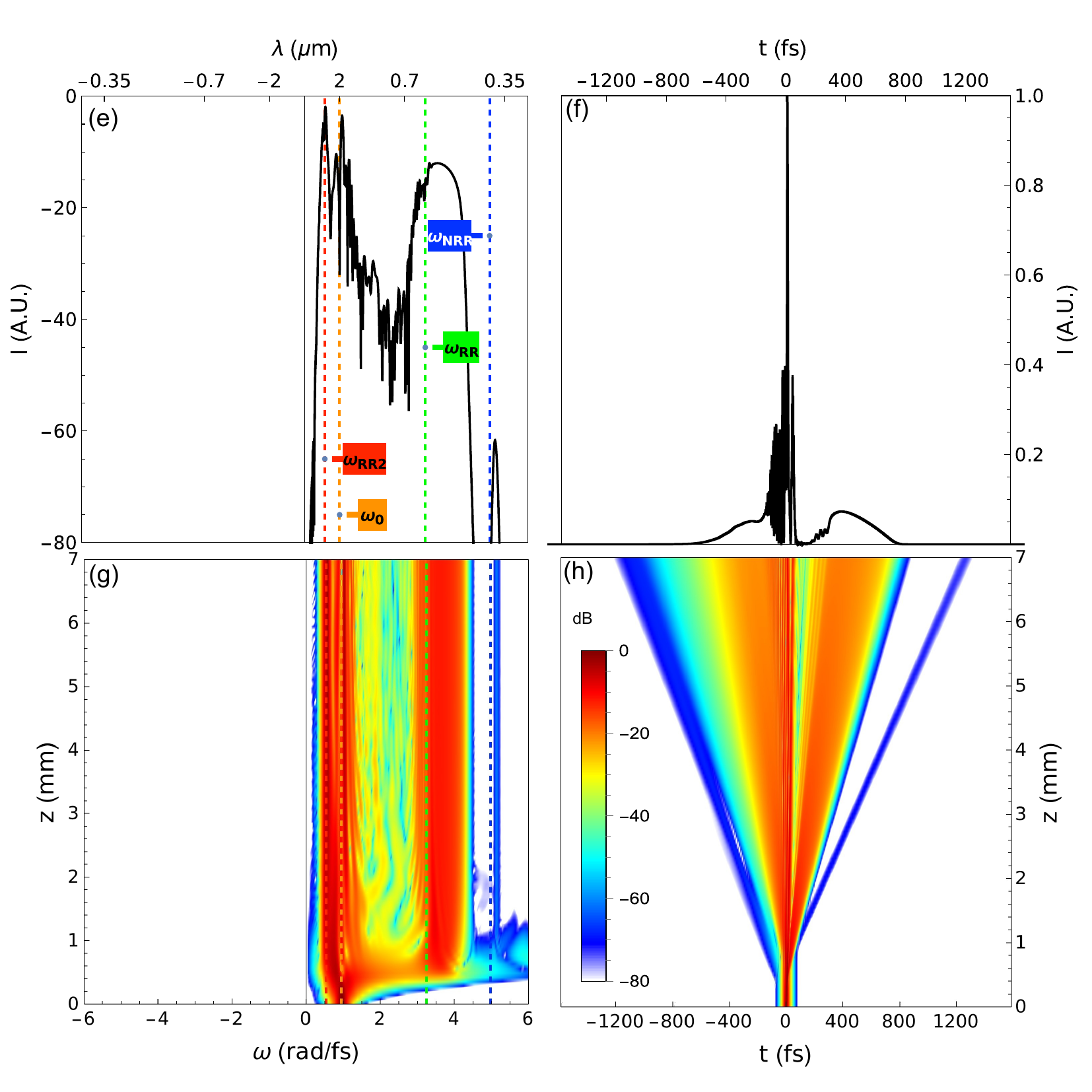}
\includegraphics[width=0.49\linewidth]{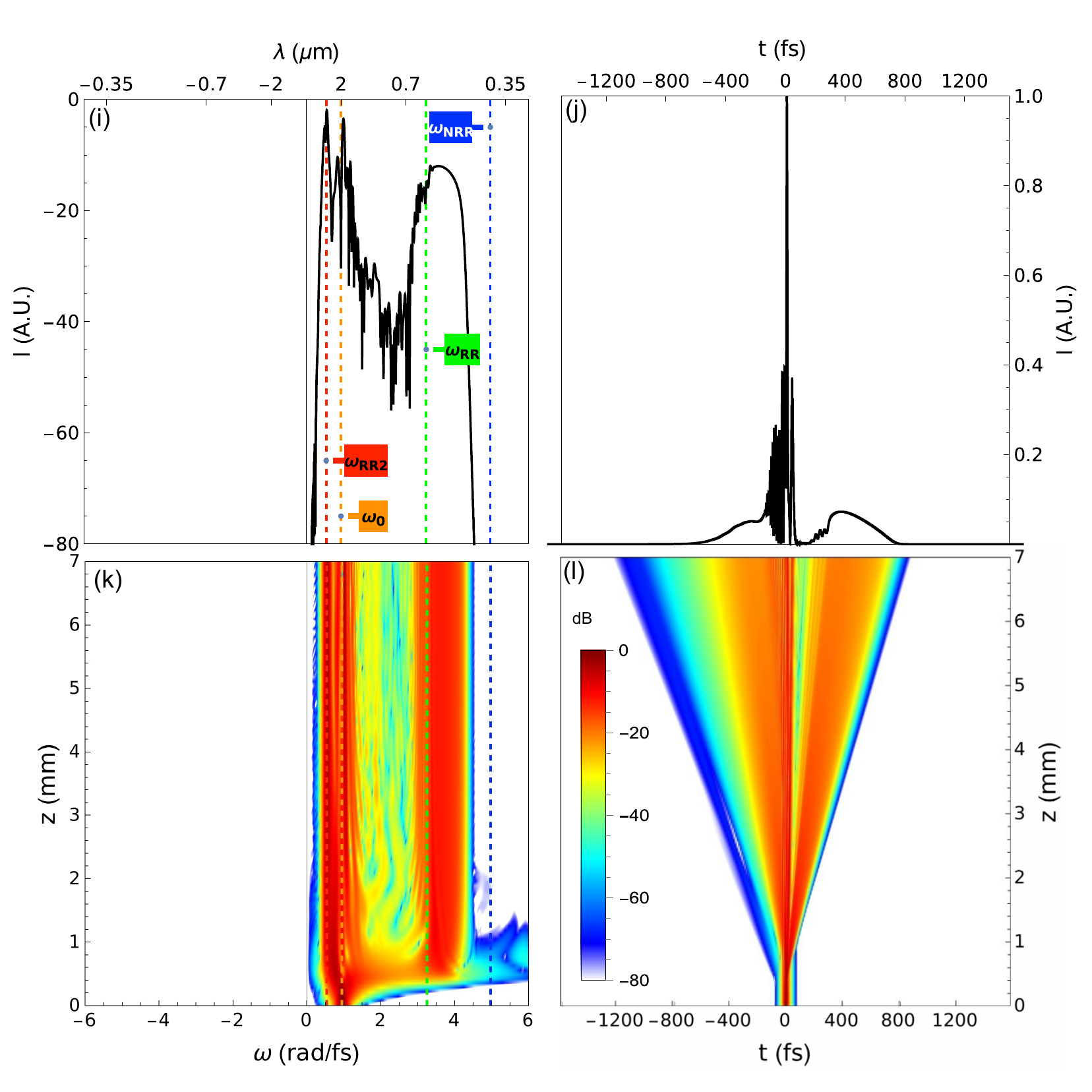}
\caption{Numerical solution to Eq. \eqref{UPPEcorrect}. Panels (a)-(d) consider all nonlinear terms, (e)-(h) with the SPM and SPM* terms, and (i)-(l) with only the SPM term. }
\label{AS_ALL}
\end{figure}

In Figure \ref{AS_ALL} (a)-(d), we show the solutions of Eq. \eqref{UPPEcorrect} for the same system as in Figure \ref{EF}, keeping all three nonlinear terms. In panels (e)-(h), we keep the SPM and SPM* terms, and in panels (i)-(l), we keep only the SPM term. The solution with all terms reproduces the physics of the solution of the electric field formalism in Eq. \eqref{UPPE}, except all information is on the positive side of the spectrum. Since the electric field is a rapidly oscillating function, the amplitudes of the two solutions are not the same, since the analytic signal is a kind of envelope of the electric field. The real part of the analytic signal is identical to the electric field in the time domain.

One of the advantages of controlling individual nonlinear terms can be seen by comparing panels (a)-(d) and (e)-(h), where we discard the THG term, or (e)-(h) and (i)-(l) where we also discard the SPM* term. Depending on the fiber properties and the pump pulse conditions, the THG signal can have a wide bandwidth to the point where it can mask other weak signals. Our numerical solutions show an example where the THG signal interferes with the NRR signal. This fact also explains why it is relatively difficult to measure the NRR signal because we cannot filter out the individual nonlinear effects in the laboratory. We can also confirm that the NRR signal originates from the SPM* term $\mathcal{E}^{*2}\mathcal{E}$, since this signal disappears only when we drop this term in panels (i)-(l). This result is consistent with those in Ref. \cite{Conforti2013}.

\section{Double analytic signal formalism}\label{secDAS}
So far, we have studied two formalisms for solving the propagation of an optical pulse: (i) using the real-valued electric field $E(z,t)$, whose spectrum contains positive and negative frequencies, and (ii) using the complex-valued analytic signal $\mathcal{E}(z,t)$, whose spectrum contains only positive frequencies. In this section, we introduce a new formalism in which we now separate the contributions of each nonlinear term into the positive and negative sides of the spectrum. We base this formalism on a new quantity we call the double analytic signal (DAS) $e(z,t)$. We obtain the correct propagation equation for the DAS, which is equivalent to the previous two formalisms. This quantity is complex---like the analytic signal---and is defined in the positive and negative parts of the spectrum---like the electric field. However, its positive and negative parts are asymmetrical, since each contains different information.

We can rewrite the UPPE for the analytic signal as
\begin{eqnarray}\label{UPPEforAS}
 &i\partial_z \mathcal{E}_\omega+\beta(\omega) \mathcal{E}_\omega + \gamma\frac{\omega}{n(\omega)} [ \mathcal{E}^3+3|\mathcal{E}|^2\mathcal{E} +3|\mathcal{E}|^2\mathcal{E}^*+\mathcal{E}^{*3}]_{\omega+} =0,
\end{eqnarray}
where, for later clarity, we restored the last term THG* that is zero due to the positive filtering procedure.

It is important to keep track of the order of the operations when working with the Fourier transform of conjugated terms. The so-called main property summarizes this fact:
\begin{eqnarray}\label{MainProperty}
(\mathcal{E}^*)_\omega &=  \mathcal{F}  \{ \mathcal{E}^{*}(t) \}= \frac{1}{2\pi} \int^\infty_{-\infty} \mathcal{E}^{*}(t) e^{-i \omega t} \dd t = \frac{1}{2\pi} \int^\infty_{-\infty} \mathcal{E}^{*}(t) e^{i (-\omega) t}\dd t\nonumber\\
&= \left[  \frac{1}{2\pi} \int^\infty_{-\infty}  \mathcal{E}^{}(t) e^{-i (-\omega) t} \dd t \right]^*= \left( \mathcal{E}_{-\omega} \right)^*,
\end{eqnarray}
where the Fourier transform of the conjugated analytic signal results in the conjugated of the Fourier transform of the analytic signal with negative frequencies only.

The DAS is a complex field with domain $\omega\in(-\infty,\infty)$, which we reduce to the analytic signal and its propagation. To do this, we partition the DAS into its positive- and negative-frequency parts $e_\omega = e_{p \omega}+ e_{n \omega}$, such that $e_{p\omega} =\Theta(\omega) e_\omega =e_{\omega +}$ and $e_{n\omega} =\Theta(-\omega) e_\omega = e_{\omega -}$, where the subindex $\pm$ indicates positive or negative frequency filtering defined by $\Theta (\pm \omega)$, as before. This step is only necessary to prove its equivalence to the analytic signal.

For this reduction, we also split the analytic signal in frequency space into two fields $p$ and $n$ such that $\mathcal{E}_\omega=\mathcal{E}_{p \omega}+\mathcal{E}_{n \omega}$ with domain $\omega\in[0,\infty)$ all fields. This separation is identical to the one performed when considering a signal formed by two well-defined and separated central frequencies. Note that both the $\mathcal{E}_{p\omega}$ and $\mathcal{E}_{n\omega}$ fields are analytic signals defined on the positive side of the spectrum and can overlap. On the other hand, in the DAS formalism, we can identify the $p$ field $e_{p\omega}$ with the positive side of the spectrum and the $n$ field $e_{n\omega}$ with the negative side.

By construction, we connect the positive-frequency part of $e_\omega$ to $\mathcal{E}_{p \omega} $, i.e., $\mathcal{E}_{p \omega}  =e_{p \omega}$. For the negative part, we reflect and conjugate $e_{n\omega}$ to transform this field into an analytic signal and connect it to $\mathcal{E}_{n \omega}$, i.e., $\mathcal{E}_{n \omega}  =(e_{n, -\omega})^*=(e_n^*)_\omega$. These assignments imply a reduction in the domain as $\omega \in[0,\infty)$ for $\mathcal{E_\omega}$ and $\omega \in(-\infty,\infty)$ for $e_\omega$, but note that in both cases the dropped negative-frequency part is zero. Thus, the relations in the time and frequency spaces are as follows:
%
\begin{alignat}{2}
& \mathcal{E}_{p \omega} =e_{p \omega}, \qquad & \mathcal{E}_p = e_p, &\label{conn1}\\
& \mathcal{E}_{n \omega} = (e_{n, -\omega})^*,\qquad & \mathcal{E}_n = e_n^*. &\label{conn2}
\end{alignat}
Note that the negative side in the DAS field $e_{n\omega}$ must be modified by reflection $R$ and conjugation ${}^*$ to be comparable with the analytic signal. Applying these operations yields
\begin{eqnarray}\label{conmutation}
\hspace{-10mm}[R(e_{n\omega}) ]^*=(e_{n,-\omega})^*= [R(\Theta(-\omega)e_\omega ) ]^*= [\Theta (\omega) e_{-\omega}]^*=[e_{-\omega+}]^*= (e_{-\omega})^*_+,
\end{eqnarray}
where the conjugation and filter operations can commute (see \ref{AppDAS}).
Note that $e_{p\omega}$ and $(e_{n,-\omega})^*$ can be reduced to analytic signals without loss of information, since their values are zero in the domain $\omega\in(-\infty,0)$. In this case, the $R$ function acts as $R(\mathcal{E}_\omega) =  \mathcal{E}_{-\omega} $, which causes a reflection in the argument for functions in frequency space. This change relates to the main property in Eq. \eqref{MainProperty}, where we show that $\mathcal{E}^*$ is composed only of negative frequencies.

Considering the connections in Eqs. \eqref{conn1} and \eqref{conn2}, our goal is to rebuild the analytic signal formalism starting from the DAS formalism, hence
\begin{equation}
 \mathcal{E}_\omega = \mathcal{E}_{p\omega} + \mathcal{E}_{n\omega} = e_{p\omega} + [e_{n,-\omega}]^*.
\end{equation}

We define the folding operation $\phi : e_\omega \in \mathbb{C} (-\infty,\infty)\rightarrow \mathcal{E}_\omega \in \mathbb{C} [0,\infty)$, such that
\begin{equation}\label{folding}
\mathcal{E}_\omega =\phi(e_\omega)=   e_{\omega+} +  [e_{-\omega}]^*_+.
\end{equation}
This operation reduces a given DAS to the corresponding analytic signal, where the equality is achieved in the domain $\omega\in [0,\infty)$.
This function helps us derive a propagation equation for the DAS and then connect it to the equation for the analytic signal. The DAS is more general than the analytic signal because it preserves information about whether a given signal originates from positive or negative frequencies. For this reason, there is no unfolding operation, i.e., the inverse relationship to Eq. \eqref{folding}.

Each nonlinear term in the DAS propagation equation must be a function of $e$ and $e^*$, generally denoted as $g(e)_\omega$. We can relate them to the corresponding nonlinear terms $\mathcal{G}(\mathcal{E})_\omega$ of the UPPE for the analytic signal with the folding operation
\begin{equation}\label{NLfolding}
\mathcal{G}(\mathcal{E})_\omega = \phi[g(e)_\omega] = g(e)_{\omega +} + [g(e)_{-\omega}]^*_{+},
\end{equation}
where $\mathcal{G}$ and $g$ are nonlinear functions of $\mathcal{E}$ and $e$, respectively, and have the same form.

Our goal is to develop a version of the UPPE for the DAS. By construction, the linear part of the equation is the same, so it has the form
\begin{equation}\label{UPPEew}
 i\partial_z e_\omega+\beta(\omega) e_\omega + \gamma\frac{\omega}{n(\omega)} p_{\text{NL},\omega} =0,
\end{equation}
where $p_{\text{NL},\omega}$ contains the nonlinear terms and is to be determined and reduced to be equal to the nonlinear part of the UPPE for the analytic signal. The nonlinear part for the analytic signal is obtained by replacing $\mathcal{E}_{}= \mathcal{E}_{p}+\mathcal{E}_{n}$ in the UPPE in Eq. \eqref{UPPEforAS}. This substitution results in a total of 20 nonlinear terms. We need to determine which ones to include in the term $p_{\text{NL},\omega}$, that we obtain the UPPE of the analytic signal \eqref{UPPEforAS} by applying the folding operation \eqref{folding} to Eq. \eqref{UPPEew}.

\begin{table}[]
\centering
\begin{tabular}{cc|cc}
\multicolumn{2}{c|}{DAS term}   & \multicolumn{2}{c}{AS term}  \\ \hline
\multicolumn{1}{c|}{\textcolor{c1}{spm$_p$}}  & $[|e_p|^2 e_p]_\omega$ & \multicolumn{1}{c|}{$[|\mathcal{E}_p|^2 (\mathcal{E}_p+\mathcal{E}^*_p)]_{\omega +}$}                          &\textcolor{c1}{SPM$_p$+SPM$^*_p$}  \\
\multicolumn{1}{c|}{\textcolor{c2}{spm$_n$}}  & $[|e_n|^2 e_n]_\omega$ & \multicolumn{1}{c|}{$[|\mathcal{E}_n|^2 (\mathcal{E}_n+\mathcal{E}^*_n)]_{\omega +}$}                          &\textcolor{c2}{SPM$_n$+SPM$^*_n$}  \\
\multicolumn{1}{c|}{\textcolor{c3}{xpm$_p$}}  & $[|e_p|^2 e_n]_\omega$ & \multicolumn{1}{c|}{$[|\mathcal{E}_p|^2 (\mathcal{E}_n+\mathcal{E}^*_n)]_{\omega +}$}                          &\textcolor{c3}{XPM$_p$+XPM$^*_p$}  \\
\multicolumn{1}{c|}{\textcolor{c4}{xpm$_n$}}  & $[|e_n|^2 e_p]_\omega$ & \multicolumn{1}{c|}{$[|\mathcal{E}_n|^2 (\mathcal{E}_p+\mathcal{E}^*_p)]_{\omega +}$}                          &\textcolor{c4}{XPM$_n$+XPM$^*_n$}  \\
\multicolumn{1}{c|}{\textcolor{c5}{thg$_p$}}  & $[e^3_p]_\omega$       & \multicolumn{1}{c|}{$[\mathcal{E}^3_p  + \mathcal{E}^{*3}_p]_{\omega +}$}                                      &\textcolor{c5}{THG$_p$+THG$^*_p$}  \\
\multicolumn{1}{c|}{\textcolor{c6}{thg$_n$}}  & $[e^3_n]_\omega$       & \multicolumn{1}{c|}{$[\mathcal{E}^3_n  + \mathcal{E}^{*3}_n]_{\omega +}$}                                      &\textcolor{c6}{THG$_n$+THG$^*_n$}  \\
\multicolumn{1}{c|}{\textcolor{c7}{sfg$_p$}}  & $[e^2_p e_n]_\omega$   & \multicolumn{1}{c|}{$[\mathcal{E}^{2}_p \mathcal{E}^{*}_n +\mathcal{E}^{*2}_p \mathcal{E}^{}_n ]_{\omega +}$}  &\textcolor{c9}{DFG$_p$+DFG$^*_p$}  \\
\multicolumn{1}{c|}{\textcolor{c8}{sfg$_n$}}  & $[e^2_n e_p]_\omega$   & \multicolumn{1}{c|}{$[\mathcal{E}^{2}_n \mathcal{E}^{*}_p +\mathcal{E}^{*2}_n \mathcal{E}^{*}_p ]_{\omega +}$} &\textcolor{c10}{DFG$_n$+DFG$^*_n$} \\
\multicolumn{1}{c|}{\textcolor{c9}{dfg$_p$}}  & $[e^2_p e^*_n]_\omega$ & \multicolumn{1}{c|}{$[\mathcal{E}^{2}_p \mathcal{E}^{}_n +\mathcal{E}^{*2}_p \mathcal{E}^{*}_n ]_{\omega +}$}  &\textcolor{c7}{SFG$_p$+SFG$^*_p$}  \\
\multicolumn{1}{c|}{\textcolor{c10}{dfg$_n$}} & $[e^2_n e^*_p]_\omega$ & \multicolumn{1}{c|}{$[\mathcal{E}^{2}_n \mathcal{E}^{}_p +\mathcal{E}^{*2}_n \mathcal{E}^{*}_p ]_{\omega +}$}  &\textcolor{c8}{SFG$_n$+SFG$^*_n$}
\end{tabular}
\caption{Equality of the nonlinear terms in the DAS formalism (lowercase) with those in the analytic signal formalism (uppercase).}
\label{tableterms}
\end{table}

For example, the spm$_p$ term $|e_p|^2 e_p$ in the DAS equation:
\begin{eqnarray}\label{example}
\phi [|e_p|^2 e_p]_\omega  &=    [|e_p|^2 e_p]_{\omega+} +  \{[|e_p|^2 e_p]_{-\omega} \}^*_+ = \left[|\mathcal{E}_p|^2 \mathcal{E}_p \right]_{\omega +} + \{ [ |\mathcal{E}_p|^2 \mathcal{E}_p ]_{-\omega}  \}^*_+\nonumber\\
&= \left[|\mathcal{E}_p|^2 \mathcal{E}_p \right]_{\omega +} + \{[ |\mathcal{E}_p|^2 \mathcal{E}_p ]^*_{}  \}_{\omega +} = \left[|\mathcal{E}_p|^2 (\mathcal{E}_p  + \mathcal{E}^*_p )\right]_{\omega +},
\end{eqnarray}
gives the SPM$_p$ and SPM${}^*_p$ terms. These two terms are part of what is originally the SPM term $\mathcal{E}^2 \mathcal{E}^*$. We summarize the results for all other nonlinear terms in Table \ref{tableterms}, where each matching color corresponds to the terms obtained by the folding operation for each DAS term. We prove all these equations in \ref{AppDAS}. We distinguish the nonlinear terms of the DAS and the analytic signal by using lowercase and uppercase letters, respectively. Not all terms are simply related to the same type of term (e.g., spm$_{p,n}$ to SPM$_{p,n}$), but some are related to different types (e.g., sfg$_{p,n}$ to DFG$_{p,n}$). The acronyms in Table \ref{tableterms} are common in nonlinear optics and are explained in \ref{AppDAS}.

We recover the 20 nonlinear terms of the analytic signal equation with only 10 terms of the DAS equation \footnote{The other 10 terms also give the correct signals, but with opposite frequency.}. These 10 terms are reduced to only two terms (spm $e^2e^*$ and thg $e^3$) when we consider the relationship $e_\omega =e_{p\omega}+e_{n\omega}$. Finally, the UPPE for the DAS is
\begin{equation}\label{UPPEDAS}
	i\partial_z e_\omega+\beta(\omega) e_\omega + \gamma\frac{\omega}{n(\omega)} [e^3+3|e|^2e]_\omega =0.
\end{equation}
Unlike the analytic signal, the DAS formalism has no positive filtering, since we have extended the frequency domain to include both positive and negative frequencies. From the numerical perspective, the final spectrum may exhibit resonant phenomena on the negative-frequency side when solving the propagation equation.

In fact, if we consider only the spm term (without thg) in Eq. \eqref{UPPEDAS}, the NRR signal appears on the negative side of the spectrum at the same value as the SPM* produces the NRR in the analytic signal formalism, but with a negative sign. From this fact we conclude that the two SPM and SPM* contain information about the negative-frequency resonant wave, and there is an equivalence between both formalisms. It also explains why there are no spm* and thg* terms in this new formalism: the standard spm term already contains the physics needed to generate negative-frequency signals. Moreover, this new DAS formalism was motivated by the second way to resolve the ill-defined Eq. \eqref{UPPEincomplete} by extending the domain of our field. Therefore, the UPPE for the DAS in Eq. \eqref{UPPEDAS} has the same form as the ill-defined version for the analytic signal in Eq. \eqref{UPPEincomplete}.

\begin{figure}
	\centering
	\includegraphics[width=0.49\linewidth]{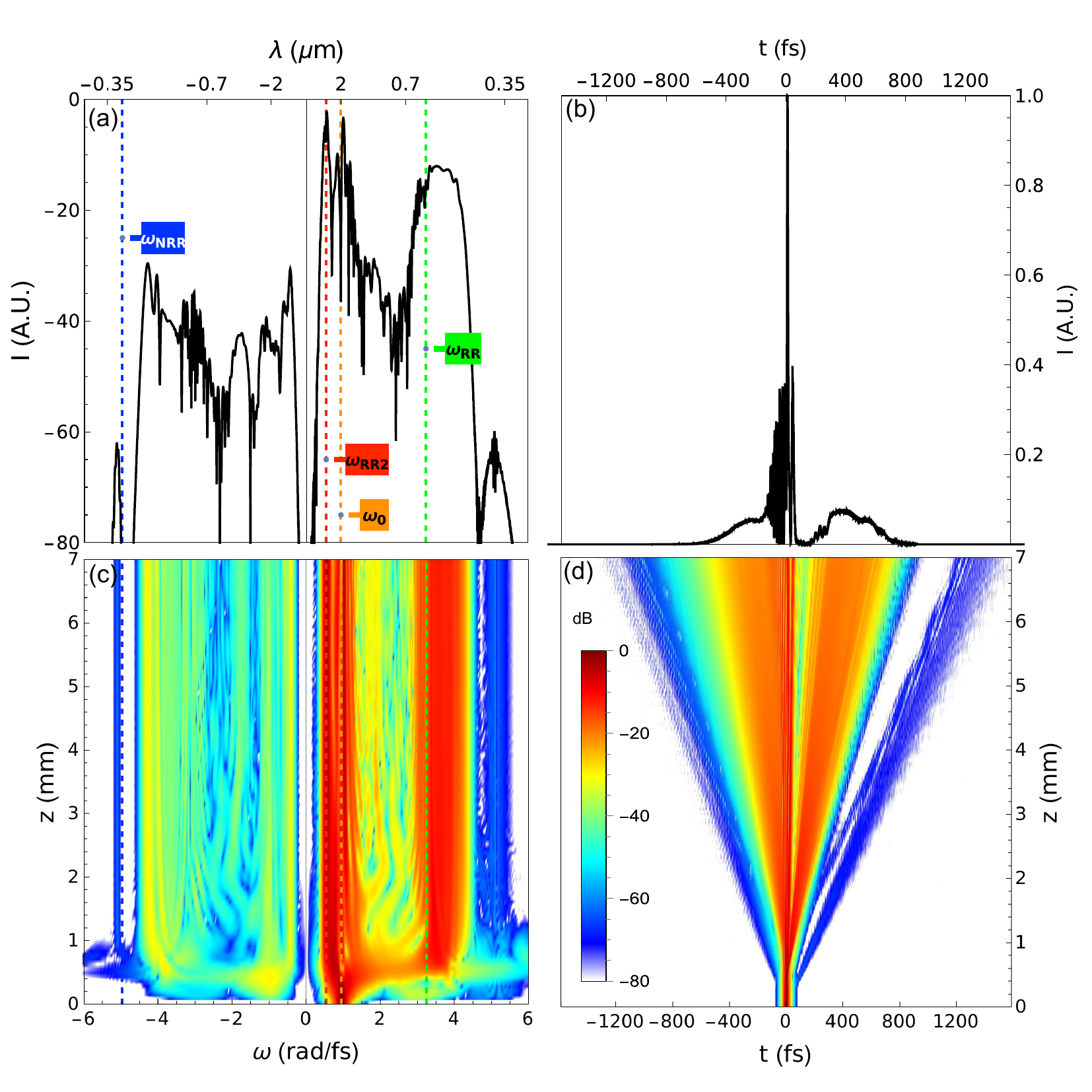}
	\includegraphics[width=0.49\linewidth]{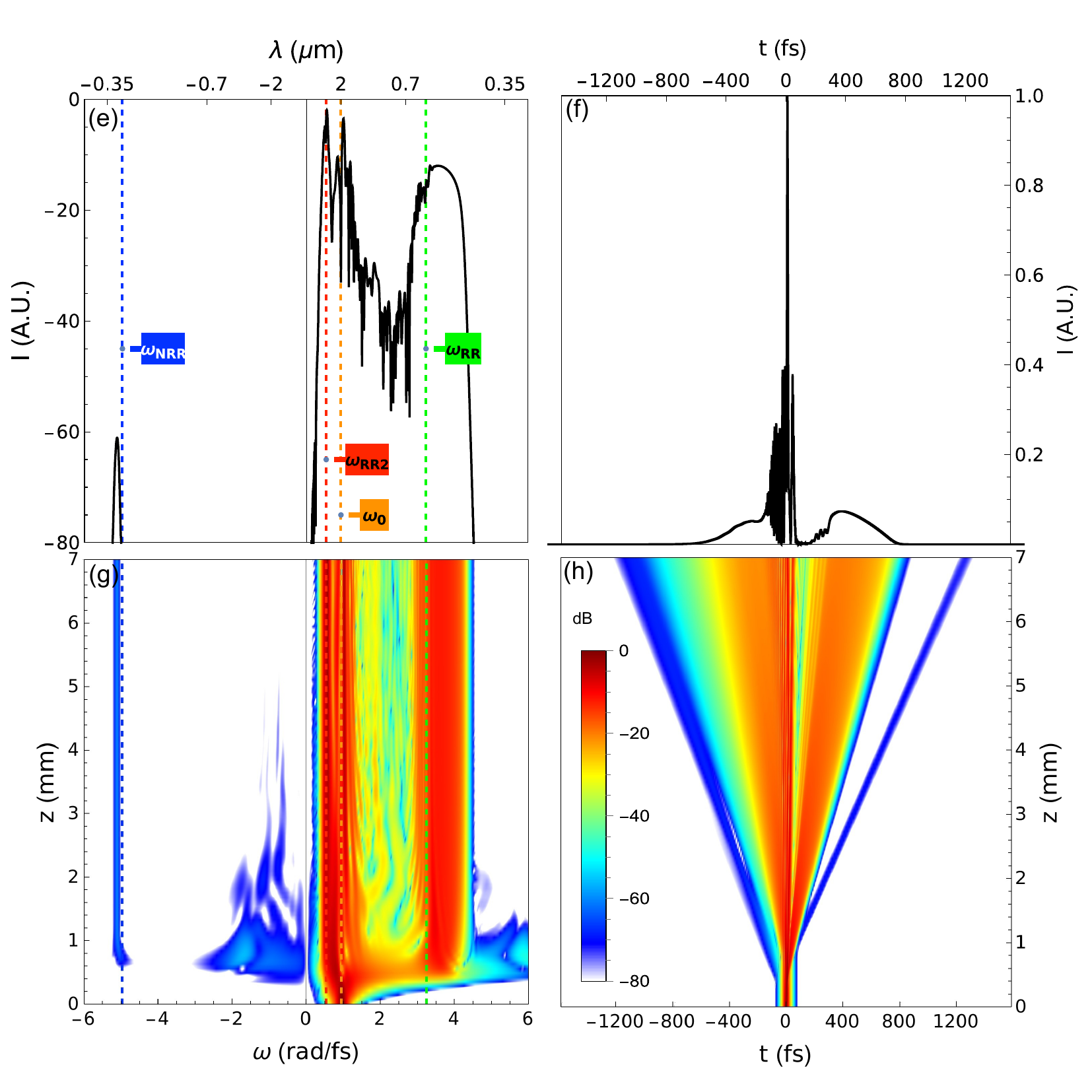}
	\caption{Numerical solution to Eq. \eqref{UPPEDAS} in the DAS formalism. In (a)-(d), we have both SPM and THG terms, and in (e)-(h), only SPM.}
	\label{DAS_ALL}
\end{figure}

Figure \ref{DAS_ALL}, panels (a)-(d) are solutions of the complete propagation equation for the DAS \eqref{UPPEDAS}. Part of the spectrum appears in the $\omega<0$ region but it is not identical to the $\omega>0$ side. All the expected signals are present,  the ones originated from negative frequencies appear in $\omega<0$. In panels (e)-(h), we keep only the spm term, and the NRR signal is kept in $\omega<0$ (besides the usual RR in $\omega>0$). It may seem to be a disadvantage that the DAS formalism only splits the nonlinear part into two terms and not into three as in the analytic signal. However, the SPM* signals are still distinguishable by their negative frequency values. In fact, now they are more visible.

We also find that the temporal solutions are identical to the analytic signal formalism. This result is unsurprising as not surprising, since we are working with the same information describing the same physical phenomena, but arranged differently.

In Figure \ref{qDAS}, we also show the DAS formalism in the dispersion relation: It has no matching condition in the $\omega'<0$ region, since the formalism does not include conjugated terms. Thus, the physics of the radiation generated by the conservation of comoving energy is contained in the I and II quadrants, as shown in the figure. In fact, this separation brings our new DAS formalism closer to the usual way of solving the Gross-Pitaevskii equation in Bose-Einstein condensates where we keep the signals from quadrants I and II \cite{Isoard2020,Giacomelli2021,Berti2023}. There, the method of mixing positive and negative frequencies involves the use of the spinor. We believe that this connection can be further explored.

\section{Conclusions}\label{secCON}
Negative frequencies occur as part of general wave phenomena and lead to experimentally verified physical signals. The sign of the frequency can be negative in cases where the emitter or the detector exceed the speed of the waves, the so-called ``supersonic'' condition. We are physically motivated to consider that the detector reads only positive frequencies, and frequencies in other frames can be negative. However, the correct description should be equivalent in all frames. The concept of negative frequencies has recently become established in optics. We discussed why negative frequencies are often missing in the propagation equations for envelope formalisms and, in particular, for the analytic signal. We showed that they are included in the electric field formalism, although this fact is unclear.

We have established how to properly include negative frequencies in the pulse propagation equation for the analytic signal (AS) and envelope equations for the electric field (EF): by keeping an extra term called conjugated self-phase modulation (SPM*) in the nonlinearity. We also note that an additional filtering procedure is necessary to maintain the positiveness of the analytic signal through propagation. The new term describes an interaction between positive- and negative-frequency signals and generates the negative-frequency resonant radiation (NRR). This NRR signal was proposed and experimentally measured by Rubino et al. in 2012 \cite{Rubino2012prl}. Shortly thereafter, it was modeled analytically and numerically with a propagation equation proposed by Conforti et al. \cite{Conforti2013}. This formalism also describes other negative-frequency effects, including the negative-frequency Hawking radiation \cite{Bermudez2016pra,Drori2019,aguero2020hawking}.

Further studies led us to formulate the double analytic signal (DAS) formalism. This mathematical object is more general than the analytic signal because it contains information about the origin of the signal. However, we showed it can be reduced to be equivalent to an analytic signal. On its own, the DAS formalism provides another way to study negative-frequency signals, originating from the interaction between positive- and negative-frequency components. The study of individual phenomena can be more precise with the effective extension of the frequency domain into an asymmetrical negative side. The comparison between the different formalisms for studying light propagation in nonlinear materials (EF vs. AS vs. DAS) shows that negative frequencies in optics are necessary to describe the system completely and conserve energy.

In Figure \ref{posneg}, we compare all the nonlinear terms for the propagation equation in each of the three different formalisms. For the electric field, there is a single nonlinear term $E^3$ that includes all nonlinear effects; without the ability to control different parts individually, we have simultaneous positive- and negative-frequency matching (red and blue, respectively). For the analytic signal, we discard the THG* term due to the filtering procedure, and the SPM and SPM* are filtered to keep only the positive-frequency matching (red). For the DAS, we only need the spm and {thg terms to recover the correct formalism, so the spm* and thg* terms are filtered out and can be discarded. 
Both the SPM and SPM* terms combine positive and negative frequencies, and we have found that they contain the same information stored in different ways: SPM gives $\omega_\text{RR}$ and $\omega_\text{NRR}$, and SPM* gives $\omega^*_\text{RR}$ and $\omega_\text{NRR}^*$. When applying the filtering procedure in the analytic signal formalism, we keep only the positive-frequency part of these terms, so we get $\omega_\text{RR}$ and $\omega_\text{NRR}^*$ on the positive side of the spectrum. In the DAS formalism, no filter acts on the spm term, and it produces $\omega_\text{RR}$ and $\omega_\text{NRR}$ in the positive- and negative-frequency sides of the spectrum, respectively.

\begin{figure}
	\centering
	\includegraphics[width=0.49\linewidth]{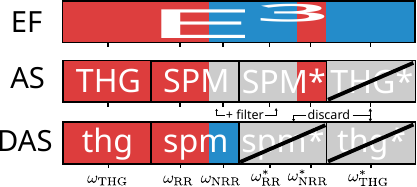}
	\caption{Comparison of the frequency content for each nonlinear term in the three formalisms. Positive and negative frequencies are colored red and blue, respectively.}
	\label{posneg}
\end{figure}
An equation for the analytic signal (Eq. \eqref{UPPEincomplete}) is often misused because it is ill-defined even though it contains all the information. The correct and complete equation for the analytic signal is Eq. \eqref{UPPEcorrect} with the filter on the nonlinear terms and the inclusion of the SPM* term. Motivated by solving the ill-definition of Eq. \eqref{UPPEincomplete}, we proposed the DAS field with extended domain to include positive and negative frequencies. We derived the correct equation for the DAS formalism given by Eq. \eqref{UPPEDAS}. Although it is identical in structure to Eq. \eqref{UPPEincomplete} for the analytic signal, they differ in the sense that we constructed the DAS field to support the entire frequency domain while retaining all nonlinear signals in an asymmetrical way. We propose implement the DAS formalism in studies of four-wave mixing, two interacting pulses, and energy conservation between different nonlinear processes.

With the DAS formalism, we clarify that the missing frequencies are not only those with negative comoving frequency, as previously thought, but also those where the matching condition includes any negative comoving frequency, even if the resulting signal has a positive comoving frequency, as shown in the negative part of Figure \ref{DAS_ALL}(a) with negative laboratory frequency and positive comoving frequency, i.e., quadrant II in Figure \ref{qDAS}. We can explore this concept further with the DAS formalism. 
Negative frequencies will become more relevant in future experiments in nonlinear optics, given the expanded experimental accessibility to conditions in the extreme nonlinear optics regime. We argue that our proposed formalism, specialized for the treatment of negative frequencies will become an essential tool for understanding the next generation of nonlinear optics experiments.

\section*{Acknowledgements}
We acknowledge funding by Conacyt Ciencia de Frontera 51458-2019 (Mexico). RAS acknowledges funding by Conacyt (Mexico) scholarship 485053.

\appendix

\section*{Appendices}

\section{Numerical solutions}\label{AppNum}
We use the Dormand-Prince method as the algorithm to numerically solve Eqs. \eqref{UPPE}, \eqref{UPPEcorrect} and \eqref{UPPEDAS}. This algorithm is a Range-Kutta method of fourth- and fifth-order where the fourth-order gives the solution and the difference between the fourth- and fifth-order is the error. We propagate each run to a value of $z=7$ mm with a step size of $0.5$ $\mu$m. We transform frequency and temporal spaces using a fast Fourier transform (FFT) with a vector size of $2^{15}$. The frequency window goes from $-8$ rad/fs to 8 rad/fs, resulting in a temporal window of 3176 fs. For the solutions of Eq. \eqref{UPPEcorrect} for the analytic signal field, we applied a $\Theta(\omega)$ function to the nonlinear part of the equation at each propagation step. For all three cases, the initial field is a soliton with central frequency $\omega_0 =0.942$ rad/fs ($\lambda = 2000$ nm), FWHM $= 15$ fs, and peak power $P_0=2\times 10^5$ W. The maximum error obtained in any solution is on the order of $10^{-4}$. We implement the code in two ways: (i) we write our own code in Wolfram Mathematica, and (ii) we use the pulse-propagation packages written in Python by Melchert et al. that are publicly available \cite{Melchert2022} as a check on our work. We modified this code to include the negative frequencies as described in this work and obtained the same results as our code, including the error.

\section{Nonlinear contributions to the DAS formalism}\label{AppDAS}

Here we present how to calculate $p_{\text{NL}}$ from the 10 nonlinear terms of the Eq. \eqref{UPPEDAS} as shown in the example in Eq. \eqref{example} for one of the spm terms.

For the self-phase modulation (spm$_p$) term $|e_p|^2 e_p$:
\begin{eqnarray}
\phi [|e_p|^2 e_p]_\omega  &=    [|e_p|^2 e_p]_{\omega+} +  \{[|e_p|^2 e_p]_{-\omega} \}^*_+ = \left[|\mathcal{E}_p|^2 \mathcal{E}_p \right]_{\omega +} +  \{[ |\mathcal{E}_p|^2 \mathcal{E}_p ]_{-\omega}  \}^*_+\nonumber\\
&= \left[|\mathcal{E}_p|^2 \mathcal{E}_p \right]_{\omega +} +  \{[ |\mathcal{E}_p|^2 \mathcal{E}_p ]^*_{}  \}_{\omega +} = \left[|\mathcal{E}_p|^2 (\mathcal{E}_p  + \mathcal{E}^*_p )\right]_{\omega +},
\end{eqnarray}

For the spm$_n$ term $|e_n|^2 e_n$:
\begin{eqnarray}
\phi [|e_n|^2 e_n]_\omega  &=     [|e_n|^2 e_n]_{\omega+} +   \{[|e_n|^2 e_n]_{-\omega} \}^*_+= \left[|\mathcal{E}_n|^2 \mathcal{E}^*_n \right]_{\omega +} +  \{[ |\mathcal{E}_n|^2 \mathcal{E}^*_n ]_{-\omega}  \}^*_+\nonumber\\
&= \left[|\mathcal{E}_n|^2 \mathcal{E}^*_n \right]_{\omega +} +  \{[ |\mathcal{E}_n|^2 \mathcal{E}^*_n ]^*_{}  \}_{\omega +}= \left[|\mathcal{E}_n|^2 (\mathcal{E}^*_n  + \mathcal{E}_n )\right]_{\omega +}
\end{eqnarray}

For the cross-phase modulation (xpm$_p$) term $|e_p|^2 e_n$:
\begin{eqnarray}
\phi [|e_p|^2 e_n]_\omega  &=    [|e_p|^2 e_n]_{\omega+} +   \{[|e_p|^2 e_n]_{-\omega} \}^*_+= \left[|\mathcal{E}_p|^2 \mathcal{E}^*_n \right]_{\omega +} +  \{[ |\mathcal{E}_p|^2 \mathcal{E}^*_n ]_{-\omega}  \}^*_+\nonumber\\
&= \left[|\mathcal{E}_p|^2 \mathcal{E}^*_n \right]_{\omega +} +  \{[ |\mathcal{E}_p|^2 \mathcal{E}^*_n ]^*_{} \}_{\omega +}= \left[|\mathcal{E}_p|^2 (\mathcal{E}^*_n  + \mathcal{E}_n )\right]_{\omega +}
\end{eqnarray}

For the xpm$_n$ term $|e_n|^2 e_p$:
\begin{eqnarray}
\phi [|e_n|^2 e_p]_\omega &=     [|e_n|^2 e_p]_{\omega+} +   \{[|e_n|^2 e_p]_{-\omega} \}^*_+= \left[|\mathcal{E}_n|^2 \mathcal{E}_p \right]_{\omega +} +  \{[ |\mathcal{E}_n|^2 \mathcal{E}_p ]_{-\omega}  \}^*_+\nonumber\\
&= \left[|\mathcal{E}_n|^2 \mathcal{E}_p \right]_{\omega +} +  \{[ |\mathcal{E}_n|^2 \mathcal{E}_p ]^*_{}  \}_{\omega +}= \left[|\mathcal{E}_n|^2 (\mathcal{E}_p  + \mathcal{E}^*_p )\right]_{\omega +}
\end{eqnarray}

For the third-harmonic generation (thg$_p$) term $e_p^3$:
\begin{eqnarray}
\phi [e_p^3]_\omega  &=   [e_p^3]_{\omega+} +   \{[e_p^3]_{-\omega} \}^*_+= \left[\mathcal{E}_p^3  \right]_{\omega +} +  \{[ \mathcal{E}_p^3  ]_{-\omega}  \}^*_+\nonumber\\
&= \left[\mathcal{E}_p^3  \right]_{\omega +} +  \{[ \mathcal{E}_p^3 ]^*_{}  \}_{\omega +}= \left[\mathcal{E}_p^3 + \mathcal{E}_p^{*3} \right]_{\omega +}
\end{eqnarray}

For the thg$_n$ term $e_n^3$:
\begin{eqnarray}
\phi [e_n^3]_\omega  &=  [e_n^3]_{\omega+} +   \{[e_n^3]_{-\omega} \}^*_+= \left[\mathcal{E}_n^{*3}  \right]_{\omega +} +  \{[ \mathcal{E}_n^{*3}  ]_{-\omega}  \}^*_+\nonumber\\
&= \left[\mathcal{E}_n^{*3}  \right]_{\omega +} +  \{[ \mathcal{E}_n^{*3} ]^*_{}  \}_{\omega +}= \left[\mathcal{E}_n^{*3} + \mathcal{E}_n^{*3} \right]_{\omega +}
\end{eqnarray}

For the sum-frequency generation (sfg$_p$) term $e_p^2 e_n$:
\begin{eqnarray}
\phi [e_p^2 e_n]_\omega  &=   [e_p^2 e_n]_{\omega+} +   \{[e_p^2 e_n]_{-\omega} \}^*_+= \left[\mathcal{E}_p^2 \mathcal{E}^*_n \right]_{\omega +} +  \{[ \mathcal{E}_p^2 \mathcal{E}^*_n ]_{-\omega}  \}^*_+\nonumber\\
&= \left[\mathcal{E}_p^2 \mathcal{E}^*_n \right]_{\omega +} +  \{[ \mathcal{E}_p^2 \mathcal{E}^*_n ]^*_{} \}_{\omega +}= \left[\mathcal{E}_p^2 \mathcal{E}^*_n  +  \mathcal{E}_p^{*2} \mathcal{E}_n \right]_{\omega +}
\end{eqnarray}

For the sfg$_n$ term $e_n^2 e_p$:
\begin{eqnarray}
\phi [e_n^2 e_p]_\omega  &=     [e_n^2 e_p]_{\omega+} +   \{[e_n^2 e_p]_{-\omega} \}^*_+= \left[\mathcal{E}_n^{*2} \mathcal{E}_p \right]_{\omega +} +  \{[ \mathcal{E}_n^{*2} \mathcal{E}_p ]_{-\omega}  \}^*_+\nonumber\\
&= \left[\mathcal{E}_n^{*2} \mathcal{E}_p \right]_{\omega +} +  \{[ \mathcal{E}_n^{*2} \mathcal{E}_p ]^*_{}  \}_{\omega +}= \left[\mathcal{E}_n^{*2} \mathcal{E}_p  +  \mathcal{E}_n^{2} \mathcal{E^*}_p   \right]_{\omega +}
\end{eqnarray}

For the difference-frequency generation (dfg$_p$) term $e_p^2 e^*_n$:
\begin{eqnarray}
\phi [e_p^2 e^*_n]_\omega  &=     [e_p^2 e^*_n]_{\omega+} +   \{[e_p^2 e^*_n]_{-\omega} \}^*_+= \left[\mathcal{E}_p^2 \mathcal{E}_n \right]_{\omega +} +  \{[ \mathcal{E}_p^2 \mathcal{E}_n ]_{-\omega}  \}^*_+\nonumber\\
&= \left[\mathcal{E}_p^2 \mathcal{E}_n \right]_{\omega +} +  \{[ \mathcal{E}_p^2 \mathcal{E}_n ]^*_{}  \}_{\omega +}= \left[\mathcal{E}_p^2 \mathcal{E}_n  +  \mathcal{E}_p^{*2} \mathcal{E}^*_n \right]_{\omega +}
\end{eqnarray}

For the dfg$_n$ term $e_n^2 e^*_p$:
\begin{eqnarray}
\phi [e_n^2 e^*_p]_\omega  &=     [e_n^2 e^*_p]_{\omega+} +   \{[e_n^2 e^*_p]_{-\omega} \}^*_+= \left[\mathcal{E}_n^{*2} \mathcal{E}^*_p \right]_{\omega +} +  \{[ \mathcal{E}_n^{*2} \mathcal{E}^*_p ]_{-\omega}  \}^*_+\nonumber\\
&= \left[\mathcal{E}_n^{*2} \mathcal{E}^*_p \right]_{\omega +} +  \{[ \mathcal{E}_n^{*2} \mathcal{E}^*_p ]^*_{}  \}_{\omega +}= \left[\mathcal{E}_n^{*2} \mathcal{E}^*_p  +  \mathcal{E}_n^{2} \mathcal{E}_p   \right]_{\omega +}
\end{eqnarray}

Equation \eqref{conmutation} can be proven by expressing the subscript $+$ as the Heaviside function:
\begin{eqnarray}	
 [X_{\omega+}]^*=[\Theta(\omega)X_\omega]^*=\Theta(\omega)^*X_\omega^*=\Theta(\omega)X_\omega^*=[X_\omega^*]_+.
\end{eqnarray}

\bibliographystyle{unsrt}
\bibliography{Refs}

\end{document}